\documentclass{emulateapj}
\shorttitle{IMAGE SUBTRACTION BY CROSS-CONVOLUTION}
\shortauthors{Yuan \& Akerlof}

\def\lsim{\mathrel{\rlap{\lower4pt\hbox{\hskip1pt$\sim$}}
    \raise1pt\hbox{$<$}}}                
\def\gsim{\mathrel{\rlap{\lower4pt\hbox{\hskip1pt$\sim$}}
    \raise1pt\hbox{$>$}}}                
\begin{document}
\title{ASTRONOMICAL IMAGE SUBTRACTION BY CROSS-CONVOLUTION}

\author{
Fang Yuan\altaffilmark{1} and Carl~W. Akerlof\altaffilmark{1}}

\altaffiltext{1}{University of Michigan, Randall Laboratory of Physics, 450
        Church St., Ann Arbor, MI, 48109-1040, yuanfang@umich.edu, cakerlof@umich.edu}

\begin{abstract}
In recent years, there has been a proliferation of wide-field sky surveys to search for a variety of
transient objects. Using relatively short focal lengths, the optics of these systems produce undersampled stellar
images often marred by a variety of aberrations. As participants in such activities, we have developed a new
algorithm for image subtraction that no longer requires high quality reference images for comparison. The computational
efficiency is comparable with similar procedures currently in use. The general technique is cross-convolution: two
convolution kernels are generated to make a test image and a reference image separately transform to match as closely as possible. In analogy to the optimization technique for generating smoothing splines, the inclusion of an RMS width penalty term constrains the diffusion of stellar images. In addition, by evaluating the convolution kernels on uniformly spaced subimages across the total area, these routines can accomodate
point spread functions that vary considerably across the focal plane.
\end{abstract}

\section{Introduction}

The advent of low-noise megapixel electronic image sensors, cheap fast computers, and terabyte data storage systems has enabled searches for rare astrophysical phenomena that would otherwise be effectively undetectable. Examples include the discoveries of MACHOs, transits of extra-solar planets, and vastly greater numbers of supernovae. Common to all of these efforts is a need to repeatedly image large portions of the sky to uncover rare and subtle changes of brightness. This forces the observer to contend with images taken under less than ideal conditions such as poor weather and crowded star fields. To solve the problem of comparing images taken with different seeing conditions, a number of research groups have developed image subtraction algorithms which compensate for image blurring effects prior to image differencing. In this paper, we describe a variation of this technique which treats pairs of images in a symmetric fashion, reducing the requirements of first obtaining an ideal reference image. The software code is relatively simple and has been made freely available.

Following our initial discovery of prompt optical radiation from a gamma-ray burst in 1999, our ROTSE collaboration set out to construct a set of four identical wide-field telescopes \citep{Akerlof} to explore these phenomena more deeply. The resulting instruments, called ROTSE-III, were installed in Australia, Texas, Namibia and Turkey in 2003 and 2004. By explicitly choosing a fast ({\it f}/1.9) optical system and short focal length (850 mm), we provided the option to search for possible "orphan" GRB afterglows without unduly compromising the potential for mapping GRB optical light curves at early times. Although we always intended to use these instruments for generic astrophysical optical transient searches, it was recognized that the plate scale (3.3" per pixel) was too coarse for easy identification of a supernova embedded in a normal galaxy.

This challenge was addressed by Robert Quimby when he became a graduate student at the
University of Texas. As an undergrad at the University of California at Berkeley, he had
worked extensively with the Supernova Cosmology Project (SCP) and was quite familiar with
the SN discovery process. Quimby adapted the SCP image subtraction code \citep{SCP_ref} for
use as the basic tool for finding 30 SNe over a period of two years from observations with
the University of Texas 30\% allocation of ROTSE-IIIb time at the McDonald Observatory.
Among those discoveries are SN 2005ap \citep{Quimby_05ap} and SN 2006gy \citep{Smith_06gy}
which appear to be the intrinsically brightest SNe ever identified.

In view of the evident success of the Texas Supernova Search (TSS) \citep{Quimby_thesis}, our group at the University of Michigan probed the image subtraction problem with the goal of applying this to the considerably more extensive image data available to the entire suite of ROTSE-III telescopes. The original hope of using the SCP code was abandoned following the realization that the program would not be made freely available. We attempted to adopt the ISIS image subtraction package\footnote{http://www2.iap.fr/users/alard/package.html} but were discouraged by the initial results. The significant undersampling of ROTSE-III stellar images coupled with asymmetric point spread functions across the image plane created a severe challenge for making clean subtractions. These issues are not satisfactorily addressed by the algorithms described by \citet{Alard_Lupton} and \citet{Alard} for two reasons: (1) we do not always have the luxury of a substantially higher quality reference image and (2), the point spread functions (PSFs) are often approximately elliptical with the axes oriented at any angle in the image plane. For a variety of reasons, the ROTSE-III PSFs can vary with temperature and telescope orientation. Thus, the possibility that one can simply convolve a new image to an ideal reference image is not always viable. With this in mind, we sought to develop a more symmetric algorithm that would be robust enough to handle less pristine observations. We should emphasize that the aim of this project is primarily for the reliable identification of transients in a very large database, not precision photometry.

\section{Mathematical Method}

The basic technique for image subtraction presented by Alard and Lupton depends on finding a suitable PSF smearing kernel, $K(u, v)$ that when convoluted with the reference image, $R(x, y)$, generates a transformed image, $R^*(x, y)$, that can be compared on a pixel-by-pixel basis with a new test image of lesser quality, $T(x, y)$:

\begin{equation}
R(x, y) \otimes  K(u, v) =  R^{\ast}(x, y)
\end{equation}

\noindent The kernel, $K(u, v)$, is constructed by a linear superposition of basis functions of the form:

\begin{equation}
f_{n,p,q}(u, v) = u^{p} v^{q} e^{-(u^{2} + v^{2})/2\sigma_{n}^{2}}
\end{equation}
 
\noindent Alard and Lupton recommend a three-fold ensemble of terms with $\sigma_n$ values spanning a 9-fold range. These functions are poor choices to synthesize an elliptical PSF at an arbitrary angle with respect to the imager sensor axes although they will be satisfactory for PSFs with close to azimuthal symmetry.  The specific values for $\sigma_n$, $p$ and $q$ must be determined by {\it ad hoc} comparisons with the characteristic PSFs associated with the particular instrument in use. The set of amplitudes for the basis functions is computed by the least squares technique to minimize the pixel-by-pixel differences between $R^*$ and $T$.

From this starting point, we decided to symmetrize the Alard-Lupton procedure by creating two convolution operators so that:

\begin{equation}
R(x, y) \otimes  K_{R}(u, v) \approx T(x, y) \otimes  K_{T}(u, v)
\end{equation}

\noindent In the limit that the reference image is substantially better than the test image, the $K_R$ operator smears $R$ with the point spread function characteristic of $T$ while $K_T$ will be essentially equal to the identity operator so that its effect on $T$ will be negligible. Under these conditions, the computation becomes functionally equivalent to the procedure adopted by Alard and Lupton. However, in general, the addition of a second convolution operator injects new mathematical degrees of freedom that must be constrained. The most obvious is that $K_R$ and $K_T$ can be multiplied by an arbitrary constant without violating image convolution equality. This can be conveniently resolved by demanding that one or both kernels be flux-conserving, ie:

\begin{equation}
\sum K(u, v) = 1
\end{equation}

The second, more complex, problem arises from the diffusion of a stellar image if $K_R$ and $K_T$ approximate broad Gaussian distributions. The convoluted image equality will be maintained but the signal-to-noise ratio of the subtracted image will drastically diminish. The solution to this was found by analogy to a similar problem in the application of smoothing splines to drawing curves through data with errors. In the latter case, one could trivially create a spline curve that ran exactly through each and every data point (as long as the abscissas are distinct). Such a curve would appear very wiggily and poorly represent the trend of the data. The solution to this problem is to add a curvature penalty term to the least squares residuals so that a trade-off is reached between adequately fitting the data and inserting unnecessarily complex behavior into the smoothed interpolation. The coefficient that scales the curvature term is a measure of the stiffness of the spline. There exists an elegant method called cross-validation that determines the stiffness parameter from the standard deviation errors for each data point.

For image subtraction, the degradation of the signal-to-noise ratio is proportional to the effective number of pixels that are summed by the convolution kernel. Since each pixel has an associated variance, $\sigma_{pix}^2$, the variance for a signal diffused over $N_{pix}$ pixels will be $N_{pix}\sigma_{pix}^2$. Assuming Gaussian distributions, we can estimate $N_{pix}$ from the width of the effective stellar point spread function, $N_{pix} \sim 4\pi (\sigma_{PSF}^2 + \sigma_K^2)$, where $\sigma_{PSF}$ is the basic stellar PSF width (in pixels) and $\sigma_K$ is the diffusive width of the convolution, $K$. If $\sigma_K^2$ is evaluated by the normal formula, $\sigma_K^2 = \sum K_ir_i^2$ where $K_i$ are the kernel element amplitudes, the sum can be deceptively small when the values for $K_i$ alternate in sign. Noting that
$\langle K_i \rangle\propto \frac{1}{\sigma_K^2}$, the value for $\sigma_K^2$ can be better estimated by $\frac{1}{2\pi}\sum K_i^2r_i^4$ which equally penalizes both positive and negative contributions to the kernel elements. Although the scaling behavior of the penalty coefficient is understood in terms of the image size, $\sigma_{pix}^2$, and $\sigma_{PSF}^2$, we have not investigated whether there is an elegant way to evaluate this quantity analogously to the cross-validation technique for splines.

\section{Computational Methods}

The image preprocessing that we require is similar to that described by Alard and Lupton. Flat-fielded images are processed by SExtractor \citep{SExtract} to create object lists with precise stellar coordinates. The IDL routine\footnote{ITT Visual Information Solutions, ITT Industries, Inc.}, {\tt POLYWARP}, is used to warp the new test image to overlay stellar objects in the reference image as closely as possible. A valid pixel map is generated to avoid pixels close to the image perimeter and screen against saturated values, etc. At this point, the fundamental image subtraction code is invoked as an IDL routine which first performs image flux normalization to equalize the mean values of the two images under comparison.

The most basic choice that the user must make is the representation of the convolution kernels. We have restricted them to $n \times n$ arrays with $n$ odd. This permits a simple representation for the convolution identity operator: $K_{[n/2], [n/2]} = 1$ while all other elements of $K_{i, j}$ are zero. For the ROTSE-III images, $n = 9$ appears to provide more than adequate coverage of stellar point spread functions under the worst conditions. 

The values for the convolution kernel elements are derived from the difference image:

\begin{equation}
D(x,y) = \{R(x, y) \otimes  K_{R}(u, v)\} -  \{T(x, y) \otimes  K_{T}(u, v)\}
\end{equation}

\noindent Invoking the criterion that $\sum D(x, y)^2$ should be a minimum subject to the requirements for $K_R$ and $K_T$ generates $2(n^{2} -1)$ linear equations via the usual least squares procedure to solve for the independent coefficients for $K_R(u, v)$ and $K_T(u, v)$ after imposing the kernel unitarity constraints. As described earlier, these equations will not provide unique solutions for $K_R$ and $K_T$ because the effective width of the two convolution kernels can still be radially scaled without substantially affecting the difference image, $D$. Thus, the quantity to be minimized must include a penalty term for radially diffusing the convoluted images any further than necessary. Following from earlier remarks, this figure-of-merit function can be represented as:

\begin{equation}
Q = \sum D(x, y)^{2} + \lambda\sum(u^{2} + v^{2})^{2}\{K_{R}(u, v)^{2} + K_{T}(u, v)^{2}\}
\end{equation}

\noindent where $\lambda$ is a constant selected to balance the contributions of the two competing error terms. From the discussion given above, the value for $\lambda$ should scale as:

\begin{equation}
\lambda = 2\pi N_{image}\frac{(\sigma_{R}^{2} + \sigma_{T}^{2})}{\sigma_{PSF}^{2}}\lambda^{\prime}
\end{equation}

\noindent where $N_{image}$ is the total number of pixels in the image, $\sigma_R^2$ and $\sigma_T^2$ are the pixel amplitude variances, $\sigma_{PSF}$ is the characteristic stellar PSF width and $\lambda^{\prime}$ is a constant of order unity.

With two $9 \times 9$ convolution kernels, the number of free parameters is 160 and the size of the regression matrix becomes problematic. The main concern is that if the images are essentially featureless (ie. no stars), the matrix elements become indistinguishable and the inverse matrix will be singular. To avoid these effects as well as various other computational issues, a binary valued mask array is created to eliminate sampling around the image perimeter, saturated pixels and all featureless areas not associated with stellar objects as determined by SExtractor. This approach was quite successful: the degree of singularity of the regression matrix was determined during the inversion process using the IDL singular value decomposition routines {\tt SVDC} and {\tt SVSOL}, codes derived from \emph{Numerical Recipes in C}\citep{Num_Recipes}.

For our ROTSE project, computational efficiency is critical because we typically acquire 400 images per night with each telescope and these must be reduced in situ comfortably within 24 hours. It was easily verified that most of the image subtraction calculations were devoted to computing the convolution kernels regression matrix described above. Examination of the two-dimensional structure of the kernels showed that the amplitudes near the edges of the $9 \times 9$ arrays were always small and suggested that the representation could be significantly reduced from 81 values to 25 by assuming a mapping from a reduced number of wavelet functions. Thus, each convolution kernel was represented by a linear superposition:

\begin{displaymath}
K(u, v) = \sum A_{i} B_{i}(u, v)
\end{displaymath}

\noindent with the basis functions, $B_i$, chosen as discrete approximations to bicubic spline functions with characteristic widths of $1$, $\frac{3}{2}$ and $2$ pixels, centered as shown in Figure 1. Using this technique shrank the regression matrix from $160 \times 160$ to $48 \times 48$ with a consequent reduction in processing time of about an order of magnitude. This brought the computation throughput to values similar to what Robert Quimby had obtained using the SCP code as adapted for his Texas Supernova Search.

Most of the image subtraction code was written in IDL with the exception of the evaluation
of the regression matrix. Since this is the core of the computational burden and IDL is
not particularly efficient in handling the necessary array indexing, this portion was
coded in C and linked to the rest of the IDL programs using the IDL standard external
calling interface. A crucial detail, particularly important for the ROTSE-III
telescopes, is the variation of the stellar PSF across the image plane. To accomodate this problem,
each $2045 \times 2049$ image was subdivided into 36 sub-images of roughly equal size. The
set of 50 kernel amplitude coefficients was calculated, one by one, for each of these
sub-images. Cross-convolved images were obtained by bilinear interpolation for every pixel
using the four nearest neighbor coefficient sets. The unitarity of the convolution kernels
is guaranteed by the linearity of the interpolation method with respect to the gridded
coefficient reference values. Although this sounds somewhat complicated, the calculation
was extremely efficient.


Another significant concern is the estimation of the background sky intensity. Initially,
we relied on SExtractor to remove this background before subtraction. However, when
applying our code to images containing large galaxies such as M31 and M33, we realized
that these backgrounds are poorly estimated around the cores of bright galaxies. The
solution we adopted only removes the background difference between the two images instead
of the individual background for each image separately. After the images are scaled so that
stellar fluxes match, a sky difference map is generated by first performing pixel by pixel
subtraction. The low frequency spatial variation of this difference image is obtained by
a process similar to the one used by SExtractor. The difference image is divided into
$32 \times 32$ pixel subimages and median pixel values are recursively evaluated, subject
to the constraint that pixels with $3 \sigma$ excursions from the median are ignored.
Saturated pixels are also excluded from the median computation. The resultant slowly varying
background is subtracted from one of the input images before invoking the cross-convolution
algorithm. The remaining common non-zero background doesn't affect the estimation of
the convolution kernels and the final subtracted image will be background-free.

A comparison of the results of the cross-convolution and the single convolution algorithms is shown in
Figure 2. In the limiting case where the PSFs of the reference image are azimuthally symmetric, the
two methods should produce rather similar results. However, when that condition is not satisfied, the
cross-convolution method is more appropriate.

For anyone wishing to employ the cross-convolution technique described in this paper, the
source code can be downloaded from the URL, http://hdl.handle.net/2027.42/57484, which points to
a set of files stored in Deep Blue, the University of Michigan institutional repository. {\bf Note:
This Web site is currently inaccessible. Please contact the authors directly.}

\section{Operational Experience}

Subtraction of a $2045 \times 2049$ ROTSE image from a reference frame using the method
described above takes approximately 4 minutes with a 2.0 GHz personal computer. If the
same reference image is used multiple times, it needs to be convolved with the base
kernels just once, saving computational time. Subtractions of three typical ROTSE images
from the same reference frame takes $\sim$ 10 minutes on the same processor. It should be
noted that the memory allocation for the process, mainly for storing the base kernel
convoluted images, scales with the size of the image and number of kernel basis sets. For
our choice of 25 kernel base vectors and a $2045 \times 2049$ image size, $\sim$ 1 GB
memory is required. This is not a serious handicap for a modern desktop computer.


Since August 2007, a supernova search pipeline using this subtraction code has been
running on images taken by the ROTSE-IIIb telescope. Selected fields with nearby rich
clusters and a high density of known galaxies are monitored on a daily basis, weather
permitting, to a typical limiting magnitude of 18.5. For each field, two sets of four 60-sec
exposures (20-sec for fields with bright target galaxies) are taken with a 30-minute
cadence. Following the method developed by the Texas Supernova Search, the images for each
4-exposure epoch are co-added as are the total 8 images for the night. All three co-additions are
subtracted by the same reference image. The difference images are processed through
SExtractor to find residual objects. To reject false detections due to bad pixels, cosmic
rays, asteroids and subtraction noise, further filtering is applied. The signal-to-noise
ratio of a
candidate has to be above 5 in the nightly 8-fold sum and 2.5 for the sum of single epoch.
The positions of a candidate in 3 subtractions must match to within one pixel for detections
above 15 SNR and 1.5 pixels for those with SNR below 15. The FWHM of the candidate has to
lie within the range of one pixel and twice the median FWHM for stars in the convolved
reference image. Finally, minimum flux change cuts are applied with a lower threshold for
detections embedded in known galaxies and higher for those corresponding to stellar objects.
This later criterion is intended to suppress variable stars.

In the 5-month period to date, the pipeline has identified all 13 reported supernovae that lie within our searched
fields. One of these initially escaped but was detected following modification of the mask size
to provide better performance during bad seeing. Also due to our early inexperience, two of these
SNe in relatively bright galaxies were initially missed during hand scanning.
In addition, the pipeline detected 7 novae in the fields of M31 and M33. Two novae rather close
to the center of M31 were missed before the background evaluation problem was addressed as
described in section 3. In terms of transient recovery efficiency, both real-world and limited Monte Carlo comparisons show that our subtraction code is comparable to the modified version of the Supernova Cosmology Project search code employed by the Texas Supernova Search.

\section{Summary}

The algorithm described in this paper can be adapted for a wide variety of photometric searches for transient objects. Its performance appears to be at least as good as other codes presently in use. Since the method is designed to handle images with significantly varying quality, it should remain effective when alternative programs may fail.

\section{Acknowledgements}

The authors especially thank Robert Quimby for his advice, suggestions and encouragement for
implementing this image analysis code and validating the results. We also appreciate the contributions of
a number of ROTSE collaborators, particularly James Aretakis, Timothy McKay, Eli Rykoff and Heather Swan. This research was supported by NASA grant NNG-04WC41G. F.~Y. was supported by NASA \emph{Swift} Guest Investigator grants NNG-06GI90G and NNX-07AF02G.

\clearpage

\begin{figure}
\plotone{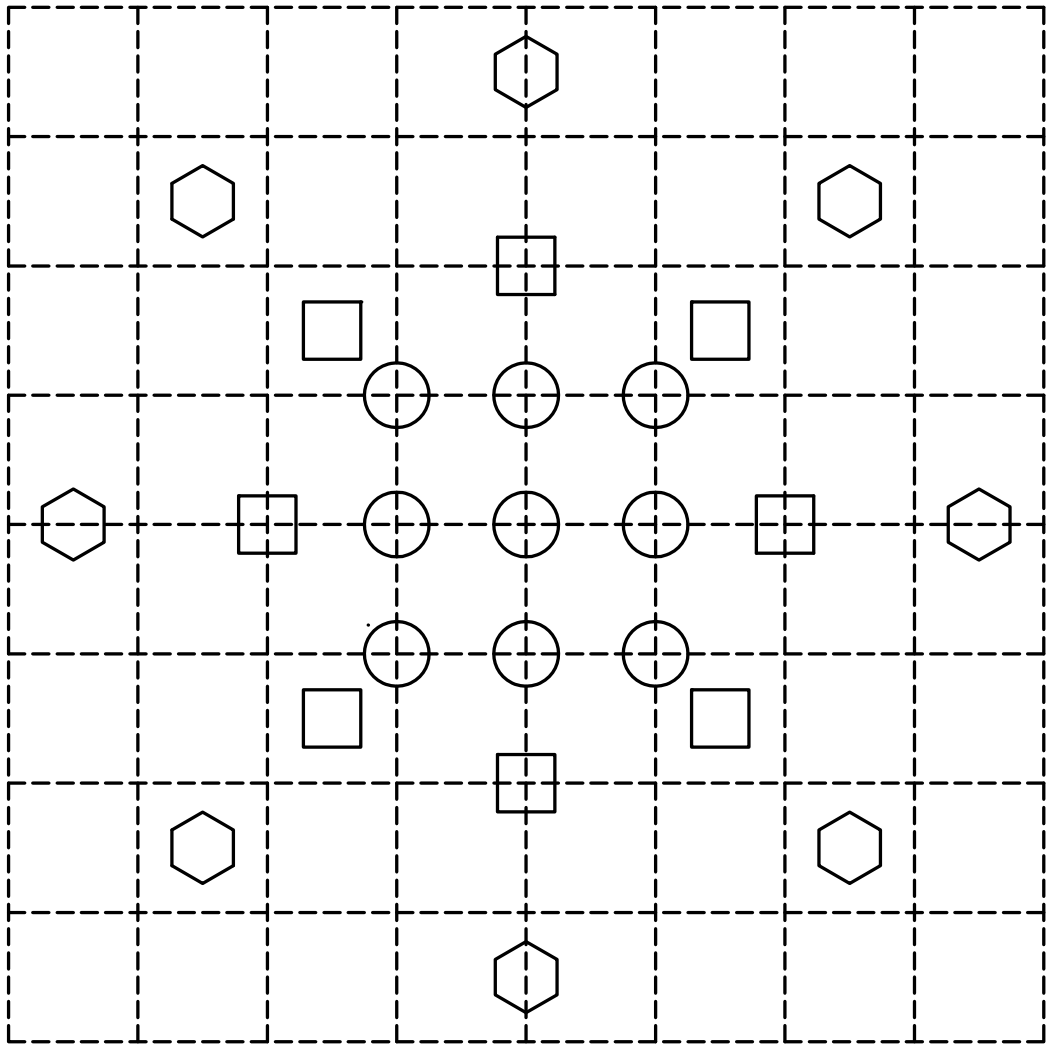}
\caption{Diagram of the location of the 25 bicubic B-splines used to construct the convolution kernels. The 9 circles, 8 squares and 8 hexagons mark the centers of the B-spline maxima with widths of $1$, $\frac{3}{2}$ and $2$ pixels respectively. The dashed lines indicate the $9 \times 9$ grid of the underlying convolution kernels.}
\end{figure}

\clearpage

\begin{figure}
\plotone{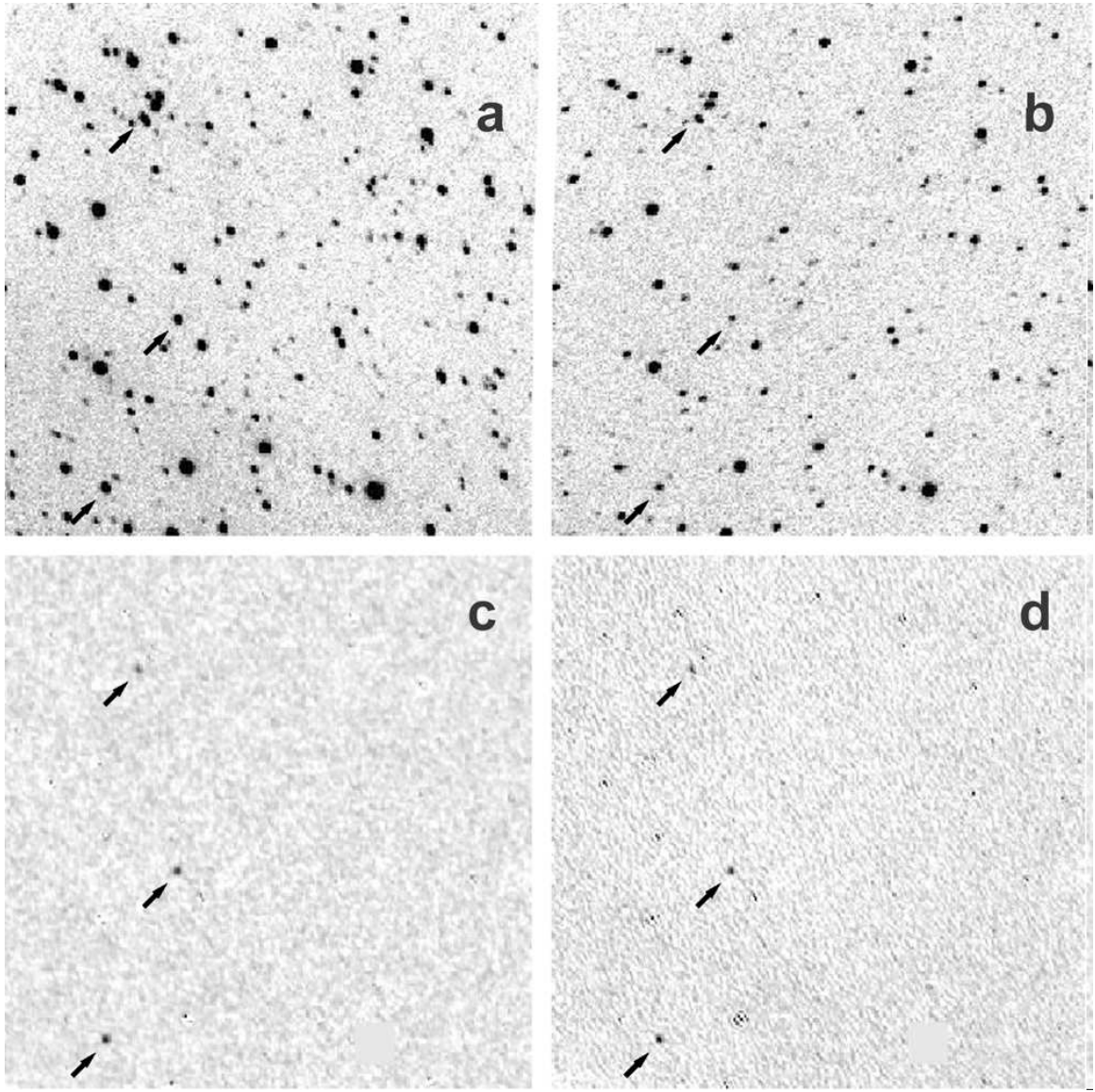}
\caption{Comparison of image subtractions using the cross-convolution method described in this paper and the single
convolution method described by Alard and Lupton and implemented in the ISIS code. The initial images were obtained by the ROTSE-IIIb telescope at McDonald Observatory. Shown here are $260 \times 260$ pixel subframes centered on
$\alpha = 16:50:02.21$, $\delta = +23:46:32.88$, covering a field of $0.235^\circ \times 0.235^\circ$. To demonstrate the results, three artificial ``variable'' stars were added to the test image (a) and the reference image (b) with PSFs appropriately matched to their respective fields. The locations are shown by black arrows. The subtracted image obtained by cross-convolution is depicted in (c) and the Alard-Lupton results are shown in (d).
The bright star near the lower right corner of the images has been replaced with a uniform gray level since neither
subtraction technique can extract useful information from saturated pixels.
 }
\end{figure}

\end{document}